\documentclass[aps,pre,showpacs,amsmath,amssymb,longbibliography]{revtex4-1}
\usepackage{amsmath,amssymb,latexsym,epsfig,graphicx,epsf,bm}
\usepackage{graphicx}
\usepackage{float}
\newcommand{\bF}{\bold F}

\newcommand{\bR}{\bold R}
\newcommand{\br}{\bold r}

\newcommand{\bRa}{{\bf R}_{\rm a}}
\newcommand{\bRb}{{\bf R}_{\rm b}}
\newcommand{\tbR}{{\tilde{\bold R}}}
\renewcommand{\tt}{\tilde{t}}
\newcommand{\tnabla}{\tilde{\nabla}}

\newcommand{\be}{\begin{equation}}
\newcommand{\ee}{\end{equation}}
\newcommand{\fig}[1]{Fig.~\ref{#1}}
\newcommand{\Fig}[1]{Figure~\ref{#1}}
\newcommand{\sect}[1]{Sec.~\ref{#1}}

\newcommand{\eq}[1]{Eq.~(\ref{#1})} 
\newcommand{\Eq}[1]{Equation~(\ref{#1})}
\newcommand{\Sex}{{S}_{\rm ex}}
\newcommand{\Tc}{{T}_{\rm conf}}

\newcommand{\Teq}{{T}_{\rm eq}}
\newcommand{\Ts}{{T}_{\rm s}}

\begin{document}
	\title{Configurational temperature in active matter. I. Lines of invariant physics in the phase diagram of the Ornstein-Uhlenbeck model}
	\date{\today}
	\author{Shibu Saw}\email{shibus@ruc.dk}
	\affiliation{\textit{Glass and Time}, IMFUFA, Department of Science and Environment, Roskilde University, P.O. Box 260, DK-4000 Roskilde, Denmark}
	\author{Lorenzo Costigliola}\affiliation{\textit{Glass and Time}, IMFUFA, Department of Science and Environment, Roskilde University, P.O. Box 260, DK-4000 Roskilde, Denmark}	
	\author{Jeppe C. Dyre}\email{dyre@ruc.dk}
	\affiliation{\textit{Glass and Time}, IMFUFA, Department of Science and Environment, Roskilde University, P.O. Box 260, DK-4000 Roskilde, Denmark}

\begin{abstract}
This paper shows that the configurational temperature of liquid-state theory, $\Tc$, defines an energy scale, which can be used for adjusting model parameters of active Ornstein-Uhlenbeck particle (AOUP) models in order to achieve approximately invariant structure and dynamics upon a density change. The required parameter changes are calculated from the variation of a single configuration's $\Tc$ for a uniform scaling of all particle coordinates. The resulting equations are justified theoretically for models involving a potential-energy function with hidden scale invariance. The validity of the procedure is illustrated by computer simulations of the Kob-Andersen binary Lennard-Jones AOUP model, demonstrating lines of approximate reduced-unit invariance of the radial distribution function and time-dependent mean-square displacement. 
\end{abstract}

\maketitle

\section{Introduction}\label{I}

Any system in thermal equilibrium has a well-defined temperature, and the temperature concept is fundamental for understanding and quantifying a system's thermodynamic and statistical-mechanical properties. In view of this it is obvious to try to generalize temperature to characterize also non-equilibrium systems. Excellent reviews of such temperatures proposed are given in Refs. \onlinecite{cas03,pow05,leu09,pug17,zha19}. Examples are the \textit{effective temperature} quantifying deviations from the fluctuation-dissipation theorem \cite{cug94,cug11,pet20} and the \textit{fictive temperature} characterizing a glass' structure in terms of the temperature at which the liquid solidified \cite{too46,scherer}. Non-equilibrium temperatures are generally motivated by the prospect of connecting properties of the non-equilibrium system to those of the same system in thermal equilibrium. That is not the background, however, of the below proposed application of liquid-state theory's \textit{configurational temperature} \cite{LLstat,rug97,pow05,him19} to active-matter models.

Active matter is an umbrella term used to describe physical systems whose building blocks can autonomously perform mechanical work. This includes fluids consisting of self-propelled particles, e.g., suspensions of swimming bacteria or animal groups, mutually-propelled particles like cytoskeletal filaments or motor proteins, cells in various contexts, bird or insect flock dynamics, etc \cite{ang11a,mar13a,bec16,ram17,sai18,das20,sha21,bow22}. Active matter is usually not time reversible. This means that a multitude of different dynamics may come into play \cite{byr22}, presenting a much richer field of study than that of ordinary time-reversible dynamics \cite{man20a}. A noted example of the features of active matter is motility-induced phase separation (MIPS), the intriguing finding that even a purely repulsive system may phase separate into high- and low-density phases \cite{vic95,das14,cat15,ram17,gey19,das20,mer20}. 

Active matter does not have states of ordinary thermal equilibrium, but there have been suggestions for mapping active-matter states to equilibrium, implying the existence of a non-equilibrium active-matter temperature. For instance, Szamel proposed an effective temperature for a single self-propelled particle \cite{sza14}, and Fodor \textit{et al.} showed \cite{fod16} that for active Ornstein-Uhlenbeck particles with a small persistence time one can identify an effective temperature from the analog of the fluctuation-dissipation theorem (see also Refs. \onlinecite{loi08,wan11,fle20}). In a parallel development, Takatori and Brady formulated a thermodynamic-type temperature for active matter based on the swim-pressure concept \cite{tak15}. 

For an ordinary non-active system in thermal equilibrium, the temperature $T$ equals $\Tc$ defined as follows \cite{rug97,pow05}. For a system of $N$ particles with collective coordinate vector $\bR\equiv (\br_1,...,\br_N)$ and potential-energy function $U(\bR)$, $k_B\Tc\equiv\langle(\nabla U)^2\rangle/\langle\nabla^2 U\rangle$ in which $k_B$ is the Boltzmann constant, $\nabla$ is the gradient operator in the $3N$-dimensional configuration space, and the sharp brackets denote canonical-ensemble averages. The proof that $\Tc=T$ in equilibrium is so simple that it deserves to be repeated here \cite{LLstat}: If $Z$ is the configuration-space partition function integral, a partial integration of 
$\langle\nabla^2 U\rangle=\int \nabla^2 U(\bR)\exp(-U(\bR)/k_BT)d\bR/Z$ leads to
$\langle\nabla^2 U\rangle=-\int \nabla U(\bR) \cdot\nabla\exp(-U(\bR)/k_BT)d\bR/Z
=\langle (\nabla U)^2\rangle/k_BT$ from which $\Tc=T$ follows.

Approaching the thermodynamic limit, the relative fluctuations of both the numerator and the denominator of $\Tc$ vanish. This means that if one defines an $\bR$-dependent configurational temperature by

\be\label{Tc_def}
k_B\Tc(\bR)
\,\equiv\,\frac{(\nabla U(\bR))^2}{\nabla^2 U(\bR)}\,,
\ee
the identity $\Tc(\bR)\cong T$ applies in the sense that deviations go to zero as $N\to\infty$. We have this limit in mind throughout and shall (mostly) ignore that $\Tc(\bR)$ fluctuates slightly for any finite system. Note that, in contrast to the standard kinetic-energy-based temperature definition, the configurational temperature is not defined for a system of free particles. Note also that configurations with $\nabla^2 U(\bR)=0$ become less likely as $N\to\infty$, so the fact that \eq{Tc_def} is not defined for such configurations is irrelevant; by the same reasoning one can ignore the existence of configurations for which $\nabla^2 U(\bR)<0$. We return briefly below to a discussion of $\Tc$ fluctuations in simulations (\fig{fig3}(b)).

Since the derivation of the configurational temperature $\Tc$ is based on the fact that the probability in the canonical ensemble of configuration $\bR$ is proportional to $\exp(-U(\bR)/k_BT)$, it would appear that $\Tc$ cannot be relevant for systems that are far from thermal equilibrium. We show in this paper, however, that $\Tc(\bR)$ may be used for tracing out lines of invariant structure and dynamics in the phase diagram of active-matter models with hidden scale invariance. This is the symmetry that the ordering of configurations according to their potential energy at a given density is maintained if these are scaled uniformly to a different density (\eq{hsi} below), a property that applies to a good approximation for the liquid and solid phases of a number of well-known potentials, including models based on the Lennard-Jones and Yukawa pair potentials \cite{IV,sch14,ing15,dyr18a}, as well as for more complicated non-pair interactions \cite{ing12b,hum15,fri19}. The companion paper (Paper II) \cite{saw23b} presents a different application of $\Tc$ to active matter by proposing that the ratio of the so-called systemic temperature \cite{dyr20} to $\Tc$ quantifies the deviation from ordinary thermal equilibrium. Both papers focus on active-matter models without orientational interactions, i.e., models based on point particles.

\section{Lines of approximately invariant physics in the phase diagram of the Kob-Andersen AOUP model}\label{II}

This section studies active Ornstein-Uhlenbeck particle (AOUP) dynamics, which has no momentum conservation and for which hydrodynamics is not taken into account. All information about the particle interactions is contained in the potential-energy function $U(\bR)$ \cite{mar13a,fil17,sha21}. In configuration space the AOUP equation of motion \cite{far15,mag15,sza15,fod16} is

\be\label{EOM_AOU}
\dot{\bR}
\,=\, \mu \bF(\bR)\,+\,\bm\eta(t)\,.
\ee
Here $\mu$ is the mobility (velocity over force) and the force vector is given by $\bF(\bR)=-\nabla U(\bR)$. The noise vector $\bm\eta(t)$ is colored according to an Ornstein-Uhlenbeck process, i.e., is a Gaussian stochastic process characterized by 

\be\label{noise}
\langle \eta_i^\alpha(t)\eta_j^\beta(t')\rangle
\,=\,\delta_{ij}\delta_{\alpha\beta}\frac{D}{\tau}\,e^{-|t-t'|/\tau}\,
\ee
in which $i$ and $j$ are particle indices, $\alpha$ and $\beta$ are spatial $xyz$ indices, and $D$ and $\tau$ are constants. We are interested in how the physics is affected if the density is changed, in particular whether approximately invariant physics can be obtained by adjusting $D$ and $\tau$ (regarding $\mu$ as a system-specific constant).

\begin{figure}[h]
	\includegraphics[width=15cm]{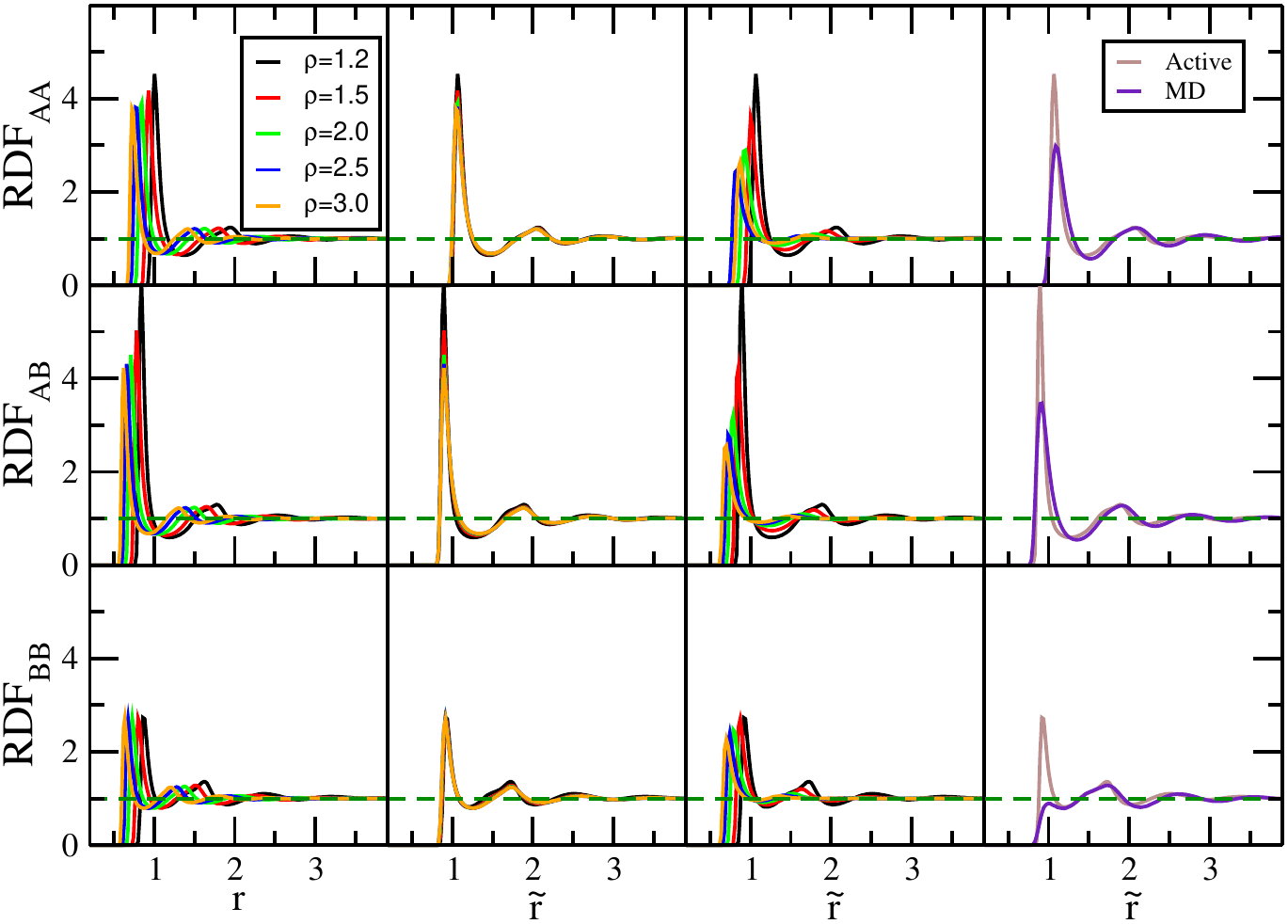}
	\caption{\label{fig1} Radial distribution functions (RDF) of the Kob-Andersen system with AOUP dynamics at densities between 1.2 and 3.0 for the model parameters $D$ and $\tau$ of Table \ref{Dtable} determined by means of \eq{Tc_def} and \eq{OUP_param2}, as described in the text. 
	The first column shows the three partial RDFs along the proposed line of invariance generated from a single configuration of the reference state point  $(\rho_0,D_0,\tau_0)=(1.2,3000,10)$, plotted as functions of the pair distance $r$. 
	The second column shows the same data as functions of the reduced pair distance $\tilde{r}\equiv\rho^{1/3}r$, revealing a good collapse except at the first peak. 
	For comparison, the third column shows data for the same values of $D$ and $\tau$ as the two previous columns at density $\rho=1.2$, while the fourth column shows AOUP data at the reference state point (brown) and standard molecular dynamics (MD) thermal equilibrium data at the density $\rho=1.2$ (indigo), evaluated at the MD temperature resulting in the same average potential energy as that of the AOUP simulation, $T_{MD}=1.57$ (the AOUP system's so-called systemic temperature \cite{dyr20}).}
\end{figure}

The dimension of $\mu$ is length squared over energy times time. Thus, if $l_0$ is a length unit, $t_0$ a time unit, and $e_0$ an energy unit, the quantity $\mu t_0e_0/l_0^2$ is dimensionless. Likewise, $Dt_0/l_0^2$ and $\tau/t_0$ are dimensionless because $D$ has dimension of a diffusion coefficient and $\tau$ of a time. It is reasonable to expect that when the density is changed, invariant physics can come about only if these three dimensionless quantities do not change -- although this criterion of course depends on the choice of units. As length unit we take the average interparticle spacing, $l_0=\rho^{-1/3}$ (in 3 dimensions). The colored-noise correlation time $\tau$ of \eq{noise} is a natural choice for the time unit, $t_0=\tau$. The idea is now to investigate the consequences of using for the energy unit the configurational temperature, i.e., of choosing $e_0=k_B\Tc$ (\sect{III} justifies this choice by reference to the isomorph theory). If the above two dimensionless quantities are to be invariant when density varies, the following must apply: $\mu\propto l_0^2/(t_0e_0)=\rho^{-2/3}/(\tau k_B\Tc)$ and $D\propto l_0^2/t_0=\rho^{-2/3}/\tau$. Since $\mu$ is assumed to be constant, this leads to $\tau\propto\rho^{-2/3}/k_B\Tc$ and $D\propto k_B\Tc$. Thus the following equations determine $D$ and $\tau$ at the different density $\rho$ from their values $D_0$ and $\tau_0$ at a reference state point of density $\rho_0$,

\begin{eqnarray}\label{OUP_param}
D&\,=\,&D_0\,\,\frac{\Tc(\rho)}{\Tc(\rho_0)}\,,\nonumber\\
\tau&\,=\,&\tau_0\left(\frac{\rho_0}{\rho}\right)^{2/3}\frac{\Tc(\rho_0)}{\Tc(\rho)}\,.
\end{eqnarray}
We note that $D$ is proportional to the configurational temperature $\Tc(\rho)$, a finding that is analogous to the equilibrium result $D\propto T$ in which $T$ is the temperature (see the next paragraph for the definition of $\Tc(\rho)$).

\begin{figure}[htbp!]
	\includegraphics[width=12cm]{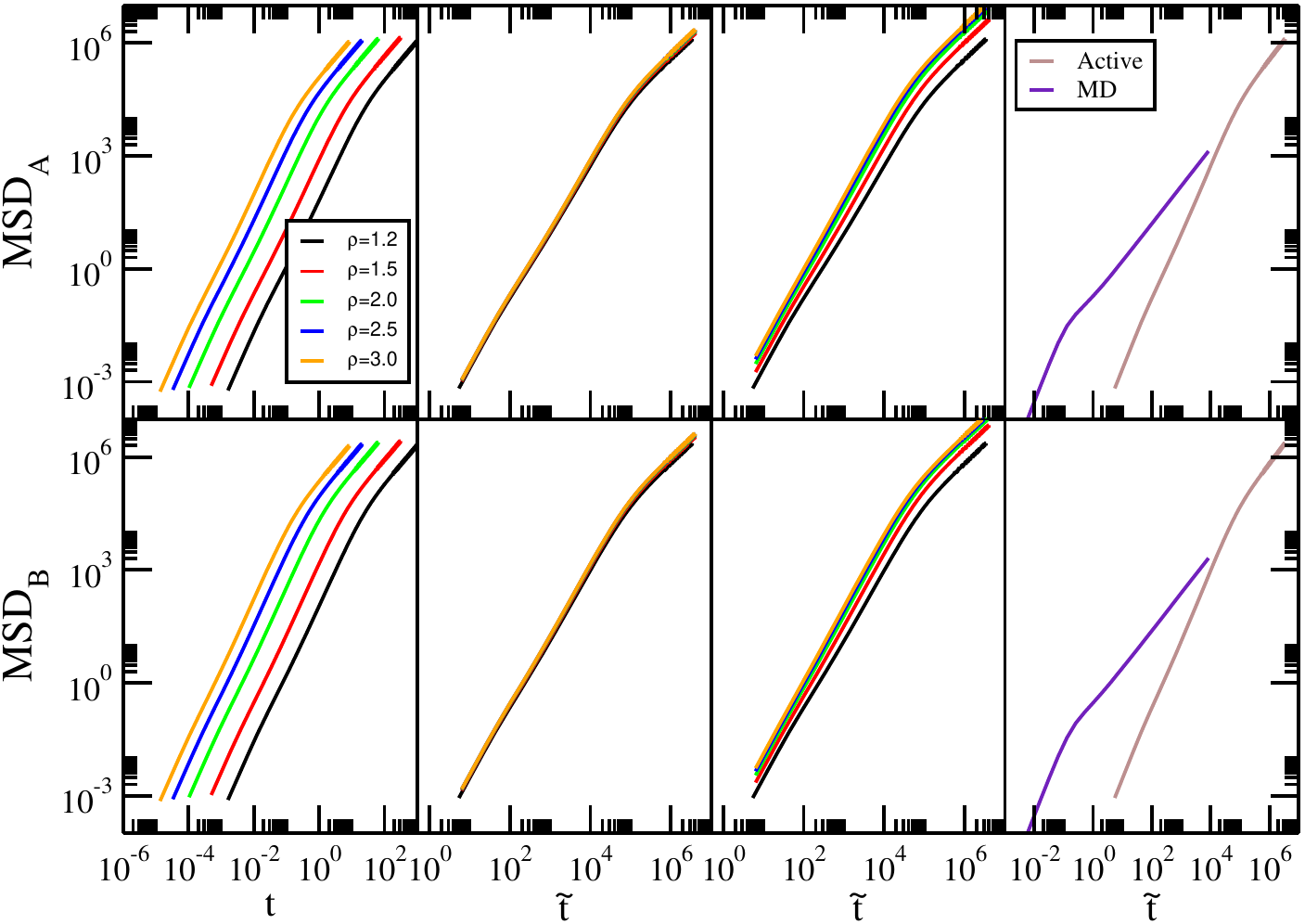}
	\caption{\label{fig2} Mean-square displacement (MSD) at the same state points as in \fig{fig1}. 
		The first column shows the MSDs of the A and B particles along the predicted line of invariance, plotted as functions of the time $t$. 
		The second column shows the same data in reduced units (defined in the text), revealing a good collapse. 
		The third column shows reduced data for the same values of $D$ and $\tau$ as the previous figures at the density $\rho=1.2$.
		The fourth column shows a comparison of the reduced MSD AOUP data at the reference state point (brown) to the standard reduced MD thermal-equilibrium MSD (indigo) at $\rho=1.2$, where the temperature as in \fig{fig1} was determined to result in the same average potential energy as that of the AOUP simulation, leading to $T_{MD}=1.57$.}
\end{figure}

As mentioned, fluctuations are small for a large system, and in that case $\Tc(\rho_0)$ may be evaluated reliably from a single configuration of a steady-state simulation, $\bR_0$: $\Tc(\rho_0)\cong\Tc(\bR_0)$. In order to find $\Tc(\rho)$ one scales $\bR_0$ uniformly to the density $\rho$ using $\bR=(\rho_0/\rho)^{1/3}\bR_0$; the configurational temperature of \eq{OUP_param} is then identified from $\Tc(\rho)\cong\Tc(\bR)$. This leads to the following recipe for calculating $D$ and $\tau$ at density $\rho$

\begin{eqnarray}\label{OUP_param2}
	D&\,=\,&D_0\,\,\frac{\Tc\left[(\rho_0/\rho)^{1/3}\bR_0\right]}{\Tc(\bR_0)}\nonumber\,,\\
	\tau&\,=\,&\tau_0\left(\frac{\rho_0}{\rho}\right)^{2/3}\frac{\Tc(\bR_0)}{\Tc\left((\rho_0/\rho)^{1/3}\bR_0\right)}\,.
\end{eqnarray}

To test the predicted invariance of structure and dynamics in reduced units when parameters vary with density according to \eq{OUP_param2}, we simulated the AOUP Kob-Andersen (KA) binary Lennard-Jones (LJ) model in three dimensions \cite{kob95}. A KA system of $10000$ particles consisting of the standard mix of two spheres, A (80\%) and B (20\%), was studied. Writing the LJ pair potential between particles of type $\alpha$ and $\beta$ as $v_{\alpha \beta}(r)=4\varepsilon_{\alpha \beta}((r/\sigma_{\alpha \beta})^{-12}-(r/\sigma_{\alpha \beta})^{-6})$ with $\alpha,\beta =A$ or $B$, the KA parameters are \cite{kob95} $\sigma_{AA}=1.0$, $\sigma_{AB}=\sigma_{BA}=0.8$, $\sigma_{BB}=0.88$, $\varepsilon_{AA}=1.0$, $\varepsilon_{AB}=\varepsilon_{BA}=1.5$, $\varepsilon_{BB}=0.5$. A shifted-force cutoff of $v_{\alpha \beta}(r)$ at $r_\textrm{cut}=2.5\sigma_{\alpha \beta}$ was used \cite{tox11a}. The simulations employed the time step $\Delta t = \Delta \tilde t / (D ~ \rho^{2/3})$ in which $\Delta \tilde t = 0.4$. At the reference density, $\rho_0=1.2$, the value of $\Delta t = 0.0001$ was used. The simulations were carried out on GPU cards; active-matter simulations used a home-made code while MD simulations used the Roskilde University Molecular Dynamics (RUMD) package \cite{RUMD}.

Table \ref{Dtable} shows the resulting values of $D$ and $\tau$ for densities ranging from 1.2 to 3.0, starting from the reference state point $(\rho,D,\tau)=(1.2,3000,10)$.

\begin{figure}[htbp!]
	\includegraphics[width=12cm]{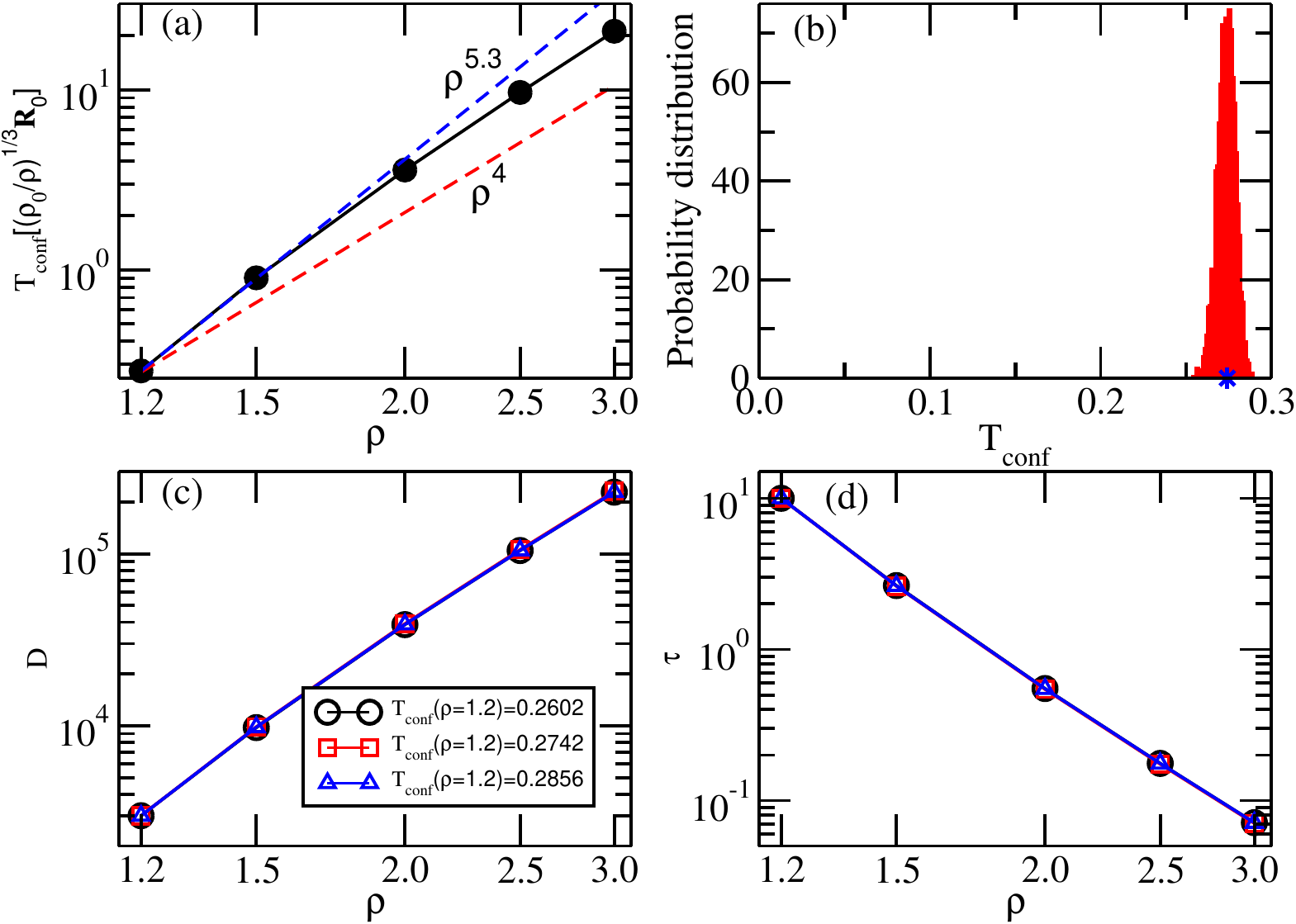}
	\caption{\label{fig3} Variation of $\Tc$, $D$, and $\tau$.
	(a) shows the configurational temperature $\Tc$ as a function of the density for the scaling configuration selected, $\bR_0$. At the highest densities, $\Tc$ is almost proportional to $\rho^4$ (red dashed line); this is where the repulsive $r^{-12}$ term of the LJ pair potential dominates the potential energy. At the lowest densities, the scaling is approximately proportional to $\rho^{5.3}$ (blue dashed line), showing that the scaling is nontrivial. 
	(b) shows the distribution of $\Tc(\bR)$ for several configurations at the reference state point for a system of $N=10000$ particles. The spread is larger than expected from a simple statistical $1/\sqrt{N}$ argument. The blue star marks $\Tc$ of the reference scaling configuration $\bR_0$.
	(c) shows how $D$ varies according to \eq{OUP_param2} for three different configurations: one is the $\bR_0$ used in \fig{fig1} and \fig{fig2} from the center of the distribution in (b) (red), the two others are from the lowest and highest ends of the distribution (black and blue). 
	(d) shows how $\tau$ varies according to \eq{OUP_param2} for the same three configurations. No significant difference are seen for the predicted parameters, meaning that $N=10000$ particles are enough for using a single configuation to determine how to scale the AOUP model parameters to obtain approximately invariant physics.} 
\end{figure}

\begin{table}[H] 
	\begin{center} 
		\begin{tabular}{|c|c|c|c|}\hline 
			$\rho$    &     $D$               &   $\tau$       &   $\Tc$    \\ \hline 
			$1.2$     &     $3000$       &   $10.000$    &     0.2742      \\ \hline 
			$1.5$     &     $9859$       &   $2.622$     &     0.9014      \\ \hline 
			$2.0$     &     $39160$      &   $0.5450$     &     3.580      \\ \hline 
			$2.5$     &     $105600$     &   $0.1741$     &     9.657      \\ \hline 
			$3.0$     &     $230800$     &   $0.0706$     &     21.10     \\ \hline 
		\end{tabular} 
		\caption{\label{Dtable} Density $\rho$ and model parameters $D$ and $\tau$ along the predicted line of invariance calculated from \eq{OUP_param2} in which $\Tc(\rho)$ is determined from a single configuration $\bR_0$ by means of \eq{Tc_def} after uniform scaling to density $\rho$.} 
	\end{center} 
\end{table}

The two left columns of \fig{fig1} show the three partial radial distribution functions (RDFs) along the predicted line of invariance shown as a function of the radial distance $r$ and of the reduced radial distance $\tilde{r}\equiv\rho^{1/3}r$, respectively. The latter shows good invariance, except that the height of the first peak is not invariant, in particular for ${\textrm{RDF}}_{\textrm{AB}}$. The third column of \fig{fig1} shows the results for the same values of $D$ and $\tau$ as previously (Table \ref{Dtable}), this time at the reference-state-point density $\rho=1.2$, in which case no invariance is observed. The fourth column compares the reference density RDFs with those of an equilibrium molecular dynamics (MD) simulation at the reference density and the temperature at which the average potential energy is equal to that of the reference-state-point AOUP simulation ($T_{MD}=1.57$), showing little resemblance. Incidentally, this ``systemic'' temperature \cite{dyr20} is quite different from the configurational temperature, $\Tc=0.27$, which corresponds to a such a deeply supercooled state for the Newtonian system that the metastable liquid cannot be equilibrated using state-of-the-art MD.

\Fig{fig2} shows the mean-square displacement (MSD) of the A and B particles as functions of time. The four columns are similar to those of \fig{fig1} with the time $t$ as the x-coordinate in the first column and the reduced time $\tilde{t}\equiv (D\rho^{2/3})t \propto t/\tau$ in the second, where the MSD is also given in reduced units, i.e., multiplied by $\rho^{2/3}$. The latter shows approximate invariance of the dynamics. It is instructive to consider the limits of short and long times. For $t\to 0$, i.e., in the ``ballistic'' regime, \eq{EOM_AOU} and \eq{noise} imply that the MSD is proportional to $(D/\tau)t^2$, while for $t\to\infty$ the MSD is proportional to $Dt$. Thus the reduced-unit short- and long-time limit MSDs are proportional to $\rho^{2/3}D\tau\tt^2$ and $\rho^{2/3}D\tau\tt$, respectively. Since \eq{OUP_param2} implies that $\rho^{2/3}D\tau$ is a constant, this means that in these limit the MSDs are proportional to $\tt^2$ and $\tt$, respectively. This is confirmed by \fig{fig2}(b). The third column of \fig{fig2} gives the reduced MSD using the predicted $D$ and $\tau$ at the reference density. The fourth column compares the reference state point MSDs to those of $T_{MD}=1.57$ MD simulation. We conclude from \fig{fig1} and \fig{fig2} that there is an approximate invariance of the reduced-unit structure and dynamics.

\Fig{fig3} investigates the robustness of the procedure. \Fig{fig3}(a) shows $\Tc$ as a function of $\rho$ in a log-log plot for selected scaling configuration $\bR_0$ used in \eq{OUP_param2}. The curve slope reveals that at high density one finds almost $\Tc(\rho)\propto\rho^4$, reflecting the dominance of the $r^{-12}$ repulsive term of the LJ pair potential (it follows from \eq{Tc_def} that $\Tc(\rho)\propto\rho^{n/3}$ for a system of $r^{-n}$ inverse power-law pair potentials). At lower densities this does not apply, however, demonstrating that the invariance of structure and dynamics is not a trivial consequence of the scale-invariant repulsive $r^{-12}$ term of the LJ pair potential. \Fig{fig3}(b) shows the distribution of configurational temperatures at the reference state point. We find a fairly broad distribution. This motivated an investigation into how much the prediction of the invariance line depends on the choice of the $\bR_0$. Figures \ref{fig3}(c) and (d) show the predictions for $D$ and $\tau$ using three different configurations in \eq{Tc_def}. The red curve is for the configuration $\bR_0$ used above that was selected from the center of the distribution in (b), the black and blue curves are for two configurations taken from the lower and higher ends of the distribution, respectively. For both $D$ and $\tau$ there is little visible difference, and we indeed find that the RDFs and MSDs are virtually indistinguishable from those of \fig{fig1} and \fig{fig2} (data not shown). Only a ratio of configurational temperatures appears in \eq{OUP_param2}, and these data suggest that a significant cancellation occurs. We conclude that, despite a relatively large spread of configurational temperatures, $N=10000$ particles are enough for \eq{OUP_param} to be used for predicting model parameters resulting in approximately invariant structure and dynamics.

\section{Theoretical justification of the procedure}\label{III}

How can the characteristic energy $k_B\Tc$ of the canonical ensemble be relevant for identifying lines of invariant physics for an active-matter system? While the energy $k_B\Tc$ \textit{per se} is not important, we argue below that the \textit{ratio} $\Tc(\rho)/\Tc(\rho_0)$ determines the \textit{ratio} of the relevant energy scales at the two densities in question. To arrive at this conclusion, we first summarize relevant aspects of the isomorph theory.

The starting point is that the KA model to a good approximation obeys the hidden-scale-invariance uniform-scaling symmetry defined \cite{sch14,dyr18a} by the following logical implication for the potential-energy function $U(\bR)$,

\be\label{hsi}
U(\bRa)<U(\bRb)\,\,\Rightarrow\,\, U(\lambda\bRa)<U(\lambda\bRb)\,.
\ee
Here $\bRa$ and $\bRb$ are configurations of the same density and $\lambda$ is a scaling parameter. Physically, \eq{hsi} expresses that the ordering of configurations at one density according to their potential energy is maintained when configurations are scaled uniformly to a different density. 

Recall that at a given thermodynamic equilibrium state point, the excess entropy $\Sex$ is defined as the entropy minus the ideal-gas entropy at the same density and temperature \cite{han13}. In the case of ordinary Newtonian mechanics, \eq{hsi} implies that structure and dynamics in reduced units (defined below) are invariant along the curves of constant excess entropy \cite{IV,sch14,dyr18a}. Such curves are termed isomorphs, and systems with isomorphs are termed R-simple. 

Isomorph invariance is exact whenever \eq{hsi} applies without exception, but this is never the case for potentials with both attractions and repulsions. Isomorph invariance is still a good approximation, however, if \eq{hsi} applies for most of the physically relevant configurations at the state points in question. Depending of course on how far the scaling parameter $\lambda$ is from unity, this is the case for the majority of  metals and van der Waals bonded systems, whereas systems with strong directional interactions like hydrogen-bonded and covalently bonded systems generally do not conform to \eq{hsi} and therefore violate isomorph-theory predictions \cite{dyr14} (ionic and dipolar systems constitute an interesting class in-between). Realistic R-simple pair-potential models include the standard Lennard-Jones model in single-component, binary, and polydisperse versions, as well as with exponents other than 6 and 12 (so-called Mie potentials), the Yukawa (screened Coulomb) pair potential \cite{sch14,vel15}, the EXP pair potential \cite{EXPI,EXPII}, effective-medium potentials describing metal \cite{fri19}, etc. We emphasize that \eq{hsi} and its consequences are not limited to pair-potential systems. For systems with inverse-power-law interactions, the isomorph theory is exact.

For R-simple systems with Newtonian dynamics, the structure and dynamics of the condensed liquid and solid phases are isomorph invariant to a good approximation when made dimensionless using as units the length $l_0$, energy $e_0$, and time $t_0$ given by (in which $m$ is the particle mass)

\be\label{red_units}
l_0=\rho^{-1/3}\,,\,\,\,e_0=k_BT\,,\,\,\,t_0=\rho^{-1/3}\sqrt{m/k_BT}\,.
\ee
Using this unit system defines the reduced-unit value of the quantity in question.

The microscopic excess-entropy function is defined \cite{sch14} by $\Sex(\bR)\equiv\Sex(\rho,U(\bR))$ in which $\Sex(\rho,U)$ is the thermodynamic excess entropy of the equilibrium state point with density $\rho$ and average potential energy $U$. Note that the function $\Sex(\bR)$ is defined for any configuration of any system, whether it is R-simple or not. It can be shown that whenever \eq{hsi} applies, $\Sex(\bR)$ depends only on the configuration's reduced coordinates, $\tbR\equiv\rho^{1/3}\bR$ \cite{sch14}. Thus inverting the relation $\Sex(\bR)=\Sex(\rho,U(\bR))$ for an R-simple system leads to 

\be\label{fundeq}
U(\bR)
\,=\,\left.U(\rho,\Sex)\right|_{\Sex=\Sex(\tbR)}
\ee  
where $U(\rho,\Sex)$ is the average potential energy of the thermodynamic equilibrium state point with density $\rho$ and excess entropy $\Sex$. For brevity, the right-hand side of \eq{fundeq} is usually written simply as $U(\rho,\Sex(\tbR))$.

The consequences of \eq{fundeq} have so far been worked out only for systems with standard time-reversible Newtonian dynamics \cite{sch14,dyr18a}. However, \eq{fundeq} follows from \eq{hsi} that has no reference to thermal equilibrium; hence \eq{fundeq} may also be applied to active-matter models with a hidden-scale-invariant potential-energy function. Note that the function $\Sex(\bR)$ still refers to the standard microcanonical ensemble according to which $\Sex(\bR)$ is basically the logarithm of the number of configurations at the same density with the same potential energy as $\bR$ \cite{sch14}.

We proceed to rewrite the AOUP equation of motion in terms of reduced variables. Writing $\bR=l_0\tbR$ and $t=t_0\tt$ in which $l_0=\rho^{-1/3}$ and $t_0=\tau$ as in \sect{II}, \eq{EOM_AOU} becomes (with $\tnabla=\rho^{-1/3}\nabla$)

\be\label{EOM_AOU_red}
\frac{l_0}{\tau}\,\dot{\tbR}
\,=\, -\mu\,\frac{1}{l_0}\,\tnabla  U(\bR)\,+\,\bm\eta(t)\,.
\ee
The reduced noise is given by $\tilde{\bm\eta}=(\tau/l_0)\bm\eta$ in terms of which \eq{noise} becomes

\be\label{noise_red}
\langle \tilde{\eta}_i^\alpha(\tt)\tilde{\eta}_j^\beta(\tt')\rangle
\,=\,\delta_{ij}\,\delta_{\alpha\beta}\,\frac{\tau\,D}{l_0^2}\,e^{-|\tt-\tt'|}\,.
\ee
\Eq{EOM_AOU_red} thus becomes

\be\label{EOM_red2}
\dot{\tbR}
\,=\, -\mu\,\frac{\tau}{l_0^2}\, \tilde{\nabla} U(\bR)\,+\,\tilde{\bm\eta}(\tt)\,.
\ee
For any configuration $\bR$ we define the ``systemic temperature'', $\Ts(\bR)$, by \cite{dyr20}

\be\label{Ts_def}
\Ts(\bR)
\,\equiv\,\left.\left(\frac{\partial U}{\partial\Sex}\right)_\rho\right|_{\Sex=\Sex(\bR)}\,.
\ee
It should be emphasized that when we below use of this concept in the context of active matter, that does \textit{not} imply an implicit mapping of the active-matter system to the ordinary thermal-equilibrium system; thus no relation between the physics of the two different cases is assumed.

In a steady-state situation the fluctuations of the systemic temperature go to zero in the thermodynamic limit, just as those of $\Tc(\bR)$. For this reason we henceforth occasionally leave out $\bR$ and write simply $\Ts$. In practice, to determine $\Ts(\bR)$ one utilizes the fact that $\Ts(\bR)$ is the equilibrium temperature $\Teq$ of the thermodynamic state point with the density of $\bR$ and excess entropy equal to $\Sex(\bR)$, implying that \cite{dyr20}

\be\label{Tseq}
\Ts(\bR)
\,=\,\Teq(\rho,\Sex(\tbR))=\Teq(\rho,U(\bR))\,.
\ee
Thus there is no need to evaluate any entropy in order to determine $\Ts$, which is simply the temperature of the thermal-equilibrium state point with same density and potential energy as the active-matter system in question.

\Eq{fundeq} implies $\tilde{\nabla} U(\bR)=\Ts\tilde{\nabla}\Sex(\tbR)$. When substituted into \eq{EOM_red2} this results in

\be\label{EOM_AOU_red2}
\dot{\tbR}
\,=\, -\mu\,\frac{\tau\Ts}{l_0^2}\,\tilde{\nabla} \Sex(\tbR)\,+\,\tilde{\bm\eta}(\tt)\,.
\ee
It follow from \eq{noise_red} and \eq{EOM_AOU_red2} that the reduced AOUP equation of motion is invariant upon a density change if $\tau D/l_0^2$ and $\tau\Ts/l_0^2$ do not vary with density. This implies $D(\rho)\propto\Ts(\rho)$ and  $\tau(\rho)\propto\rho^{-2/3}/\Ts(\rho)$ in which $\Ts(\rho)$ is short-hand notation for $\Teq(\rho,\Sex(\tbR))$, compare \eq{Tseq}. Working from the reference state point $(\rho_0,D_0,\tau_0)$, this means that the function $\Ts(\rho)$ determines how to scale $D$ and $\tau$ to ensure invariant AOUP dynamics,

\begin{eqnarray}\label{condi}
D(\rho)
\,&=&\,D(\rho_0)\,\frac{\Ts(\rho)}{\Ts(\rho_0)}\nonumber\\
\tau(\rho)
\,&=&\,\tau(\rho_0)\,\left(\frac{\rho_0}{\rho}\right)^{2/3}
\,\frac{\Ts(\rho_0)}{\Ts(\rho)}\,.
\end{eqnarray}

We next link to the configurational temperature. There is no reason to expect $\Tc=\Ts$ in out-of-equilibrium situations, and these quantities indeed differ by up to a factor of six in our simulations (Paper II \cite{saw23b} suggests using $\Ts/\Tc$ as a measure of the degree of deviation from thermal equilibrium). However, \eq{condi} still applies with $\Tc$ instead of $\Ts$ if the two temperatures are proportional in their density variation. To show that this is the case we note that \eq{fundeq} implies $\tnabla U(\bR)=\Ts\tnabla\Sex(\tbR)$ and $\tnabla^2 U(\bR)=\Ts\tnabla^2\Sex(\tbR)$, so $\Tc(\bR)=(\nabla U(\bR))^2/\nabla^2U(\bR)=(\tnabla U(\bR))^2/\tnabla^2U(\bR) =
\Ts(\tnabla\Sex(\tbR))^2/\tnabla^2\Sex(\tbR)$ \cite{dyr20}. Here we ignored the dependence of $\Ts(\bR)$ on the configuration $\bR$ at a given state point which, as mentioned above, vanishes in the thermodynamic limit. In terms of $\phi(\tbR)\equiv(\tnabla\Sex(\tbR))^2/\tnabla^2\Sex(\tbR)$ we thus have

\be\label{TsTcond}
\frac{\Tc(\bR)}{\Tc(\bR_0)} =\frac{\Ts(\rho)\,\phi(\tbR)}{\Ts(\rho_0)\,\phi(\tbR_0)}\,.
\ee
Since $\tbR=\tbR_0$ this implies $\Tc(\bR)/\Tc(\bR_0)=\Ts(\rho)/\Ts(\rho_0)$. In this way \eq{condi} leads to \eq{OUP_param}. Note that by using $\Tc$ instead of $\Ts$, one does not have to identify the equilibrium state point with the same potential energy as the active-matter state point in question. \Fig{fig1} and \fig{fig2} demonstrated good invariance along active-matter isomorphs determined by means of \eq{OUP_param}. A more accurate, but also more tedious, method for determining the active-matter isomorphs of AOUP LJ systems, which utilizes $\Ts$ directly, is discussed in the Appendix.

\begin{figure}[h]
	\includegraphics[width=7cm]{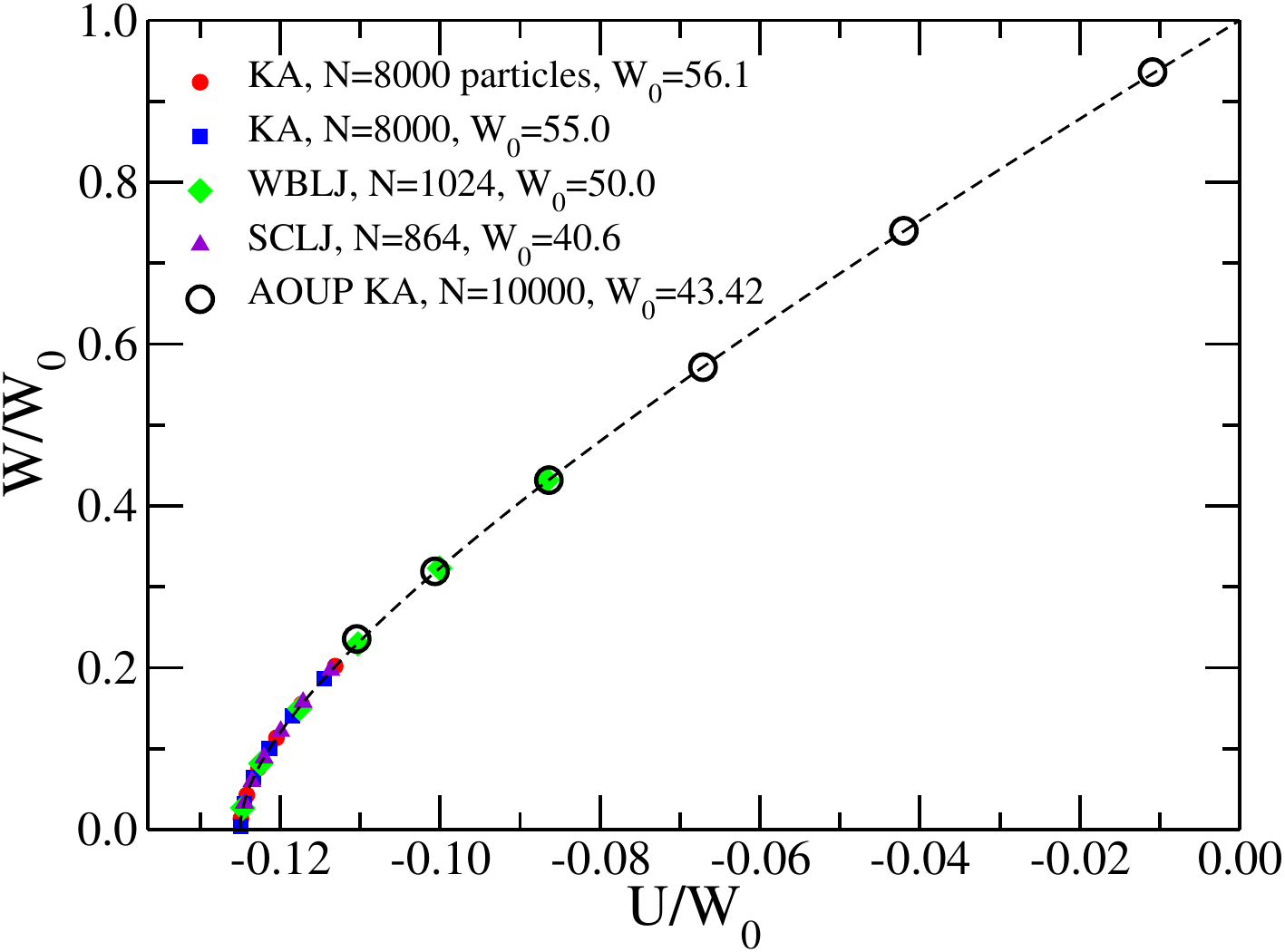}
	\caption{\label{fig4} ``Master isomorph'' (dashed curve, \eq{mi}) expressing the virial $W$ as a function of the potential energy $U$ along any isomorph of any type of LJ systems; $W_0$ is the virial at the state point of zero potential energy on the isomorph in question. The figure is Fig. 8 of Ref. \onlinecite{V} to which we have added data for the AOUP KA model (open circles). The abbreviations KA, WBLJ, SCLJ represent the Kob-Andersen system, the Wahnstrom binary LJ mixture \cite{wah91}, and the standard single-component LJ system, respectively.} 
\end{figure}

We end this section by checking a consequence of the above. It was shown in Ref. \onlinecite{V} that for LJ systems, if $W_0$ is the virial for the state point of zero potential energy on a given isomorph, the following relation between the virial $W$ and the potential energy $U$ applies along the isomorph 

\be\label{mi}
2\,\frac{W}{W_0}
\,=\,1\,+\,8\,\frac{U}{W_0}\,+\,\sqrt{1\,+\,8\,\frac{U}{W_0}}\,.
\ee
This identity is a consequence of the reduced-unit RDF isomorph invariance, which applies in the R-simple region of any single- or multicomponent LJ system (i.e., liquid or solid, but not gas) \cite{V}. Since the reduced RDF is also invariant to a good approximation along the above studied KA active-matter isomorph (\fig{fig1}), $W$ should also in this case be determined by $U$ according to \eq{mi}. This prediction is validated in \fig{fig4} that reproduces the Newtonian-dynamics equilibrium data of Ref. \onlinecite{V} to which some of our data have been added (open circles).

\section{Discussion}

The isomorph concept of equilibrium Newtonian dynamics was recently generalized to out-of-equilibrium Newtonian systems like that of a shear flow or an aging glass, leading to the introduction of the systemic temperature (\eq{Ts_def}), which allows for the identification of lines of approximately invariant structure and dynamics in the relevant out-of-equilibrium phase diagram \cite{dyr20}. The results of the present paper extend these findings by demonstrating the existence of isomorphs for active-matter systems, which in contrast to Newtonian systems are described by a dynamics that is usually not time-reversible.

The configurational temperature expression is derived from the canonical ensemble. This paper has nevertheless demonstrated the relevance of $\Tc$ for tracing out lines of approximately invariant physics in the phase diagram of Ornstein-Uhlenbeck active-matter models involving a potential-energy function that obeys hidden scale invariance. We emphasize that this application is not based on a mapping of the active-matter system to an equilibrium system.

Paper II \cite{saw23b} proposes a second application of $\Tc$ to active matter. There it is argued that the ratio of the systemic to the configurational temperature, which is unity in canonical-ensemble thermal equilibrium because $T=\Ts=\Tc$, provides a simple measure of the degree of deviation from thermal equilibrium.

\begin{acknowledgments}
	We would like to thank Thomas Voigtmann for several useful discussions and Thomas Schr{\o}der for suggesting to test the master isomorph prediction. This work was supported by the VILLUM Foundation's \textit{Matter} grant (16515).
\end{acknowledgments}

\section*{Appendix: A more accurate method for tracing out active-matter isomorphs}

Equilibrium isomorphs may be traced out by different methods, e.g., step-by-step integration of the exact equation for a configurational adiabat \cite{IV,att21}, the direct isomorph check \cite{IV}, and the recently introduced force method \cite{sch22}. These are numerical methods of varying complexity and accuracy. Likewise, there are different numerical methods for tracing out an active-matter isomorph. We describe below a more accurate alternative to the method used in Secs. II and III, a method that unfortunately is also more involved.  

The paper demonstrated how $\Tc$ can be used for tracing out active-matter isomorphs from a single configuration of the AOUP model. The result was a recipe for calculating how the model parameters are to be changed as functions of the density in order to arrive at approximately invariant structure and dynamics, \eq{OUP_param2}. This recipe provides a useful ``quick-and-dirty'' method which, since it relies on \eq{fundeq}, is exact whenever hidden scale invariance holds exactly (which is the case if $U(\bR)$ is an Euler-homogeneous function). A more accurate, but also more cumbersome, method for identifying active-matter isomorphs refers directly to the concept of systemic isomorphs. These lines in the $(\rho,\Ts)$ phase diagram are by definition the same as the ordinary isomorphs of the canonical-ensemble equilibrium $(\rho,T)$ phase diagram \cite{dyr20} (ordinary, systemic, and active-matter isomorphs are all defined as lines of constant $\Sex$ in the respective phase diagrams). 

\begin{figure}[htbp!]
	\includegraphics[width=10cm]{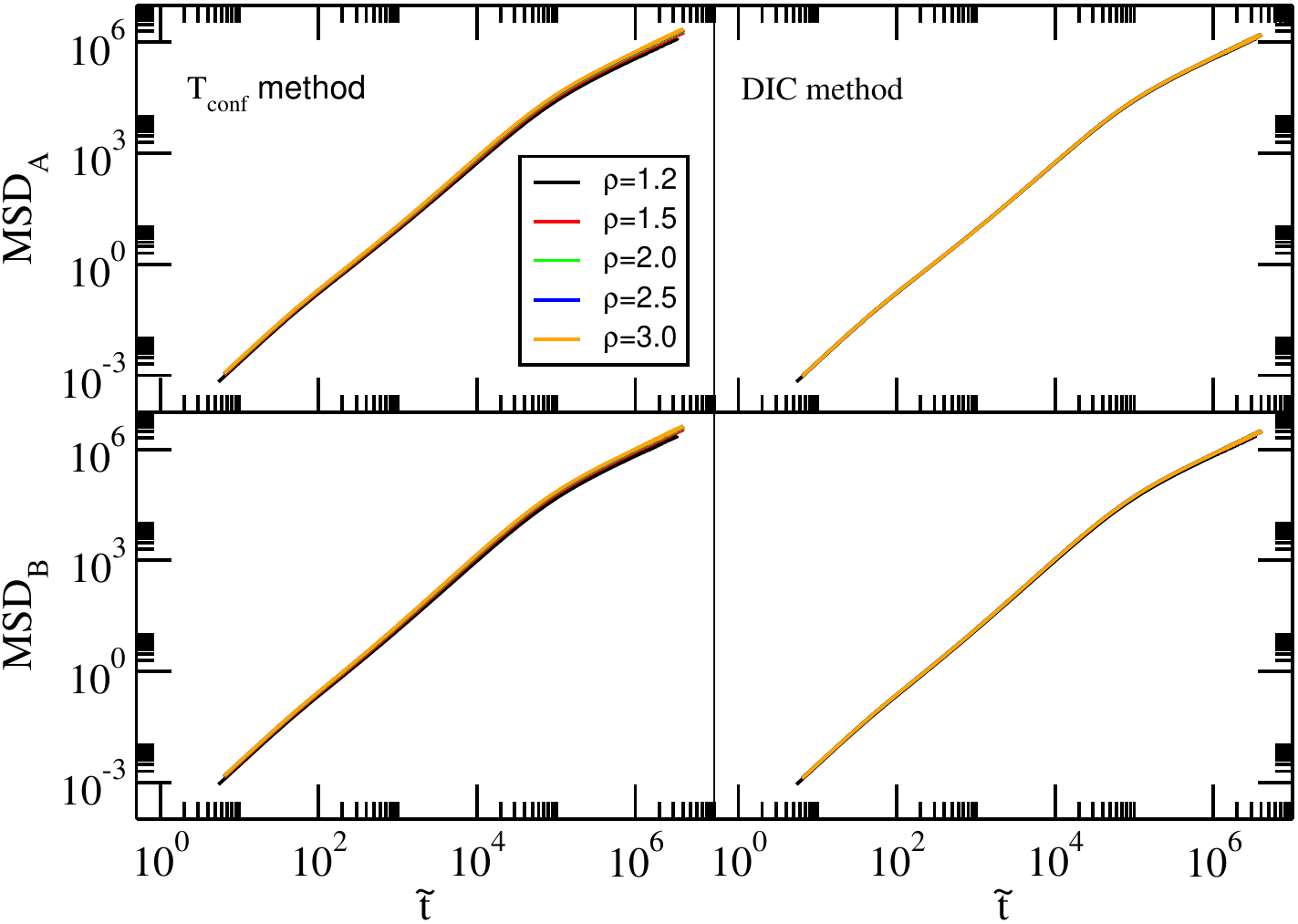}
	\caption{\label{fig5} MSD of the A and B particles as functions of reduced time along active-matter isomorphs of the AOUP KA model traced out by two different methods. The left column reproduces the data of \fig{fig2} where the isomorph was traced out by the above-developed $\Tc$ method. For comparison, the right column gives MSDs when the isomorph is traced out by the direct isomorph check (DIC) method \cite{IV} in its analytical version for LJ-type systems \cite{ing12a,boh13}, which refers directly to the fact that a systemic isomorph in the $(\rho,\Ts)$ phase diagram is identical to an equilibrium isomorph in the standard $(\rho,T)$ phase diagram. The active-matter-isomorph invariance is improved by this method.} 
\end{figure}

To test the consequence of referring directly to the systemic isomorph, we traced out the systemic isomorph of the KA system by the direct isomorph check (DIC) method \cite{IV}, which is accurate and has a simple analytical expression for LJ-type systems \cite{ing12a,boh13}. It does not make a huge difference which method is used, but there is some improvement using the DIC method. This is illustrated in \fig{fig5}, which in the left column from \fig{fig2} reproduces the reduced MSD as a function of reduced time for the A and B particles. The right column gives similar data when the model parameters are instead determined by identifying the function $\Ts(\rho)$ by utilizing the fact that this function is identical to $T(\rho)$ of the corresponding equilibrium isomorph, which may be determined by the DIC method \cite{ing12a}. We see that the MSD is more invariant in the latter case, confirming that this method is indeed more accurate.

\begin{table}[H] 
	\begin{center} 
		\begin{tabular}{|c|c|c|}\hline 
			$\rho$    &     $D_c/D_s$   &   $\tau_c/\tau_s$      \\ \hline 
			$1.2$     &     $1.000$     &   $1.000$     \\ \hline 
			$1.5$     &     $1.139$     &   $0.878$     \\ \hline 
			$2.0$     &     $1.278$     &   $0.782$     \\ \hline 
			$2.5$     &     $1.345$     &   $0.743$     \\ \hline 
			$3.0$     &     $1.382$     &   $0.723$     \\ \hline 
		\end{tabular} 
		\caption{\label{Dratiotable} Ratio of the model parameters $D$ and $\tau$ along the predicted line of invariance calculated from \eq{OUP_param2} (subscript ``c'') and from the fact that the systemic isomorph corresponds to an equilibrium isomorph (subscript ``s'').} 
	\end{center} 
\end{table}

\end{document}